\documentclass[11pt]{article}
\usepackage{fancyhdr}
\def\eqalign#1{\null\,\vcenter{\openup\jot
  \ialign{\strut\hfil$\displaystyle{##}$&$\displaystyle{{}##}$\hfil
      \crcr#1\crcr}}\,}

\usepackage{hyperref}

\def\iniz{\setcounter{equation}{0}{%
\rhead{\thepage}\lhead{{{{\small\bf\thesection:}
\small \ \SEC\ \  \tiny\today}}}}}

\let\a=\alpha \let\b=\beta    \let\d=\delta 
\let\z=\zeta  \let\h=\eta   \let\th=\vartheta  \let\l=\lambda
    \let\n=\nu    \let\x=\xi     \let\p=\pi    \let\r=\rho
\let\s=\sigma \let\t=\tau   \let\f=\varphi 
     
 \let\D=\Delta  \let\L=\Lambda

\def\V#1{{\bf#1}}
\def\*{\vskip 3mm}\def\0{\noindent}
\def\be{\begin{equation}}
\def\ee{\end{equation}}
\def\bea{\begin{eqnarray}}
\def\eea{\end{eqnarray}}

\def\BB{{\cal B}}\def\LL{{\mathcal L}}

\let\dpr=\partial

\def\Eq#1{\label{#1}}
\def\equ#1{(\ref{#1})}
\def\lis#1{\overline#1}
\def\defi{{\buildrel def\over=}}
\def\media#1{{\langle\,#1\,\rangle}}
\def\bra#1{{\langle#1|}}\def\ket#1{{|#1\rangle}}
\def\braket#1#2{\langle#1|#2\rangle}
\def\otto{\,{\kern-1.truept\leftarrow\kern-5.truept\to\kern-1.truept}\,}
\def\wt#1{{\widetilde#1}}

\def\tende#1{\,\vtop{\ialign{##\crcr\rightarrowfill\crcr
 \noalign{\kern-1pt\nointerlineskip} \hskip3.pt${\scriptstyle
 #1}$\hskip3.pt\crcr}}\,}
\def\Ie{{\it i.e.\ }}
\def\Eq#1{{\label{#1}}%
}
\newdimen\xshift \newdimen\xwidth \newdimen\yshift \newdimen\ywidth
\def\ins#1#2#3{\vbox to0pt{\kern-#2\hbox{\kern#1 #3}\vss}\nointerlineskip}

\def\eqfig#1#2#3#4#5{
\par\xwidth=#1 \xshift=\hsize \advance\xshift
by-\xwidth \divide\xshift by 2
\yshift=#2 \divide\yshift by 2%
{\hglue\xshift \vbox to #2{\vfil
#3 \includegraphics{#4.eps}
}\hfill\raise\yshift\hbox{#5}}}

\def\Bz   {\mbox{\boldmath$\zeta$}}

\def\Bx   {\mbox{\boldmath$ \xi$}}

\begin{document}

\centerline{\bf\huge Nonequilibrium stationary state}
\centerline{\bf\huge  for a damped rotator}
\*
%\centerline{\tt xxx yyyy zzzzz }
\centerline{\tt G.Gallavotti, A.Iacobucci, S.Olla}
%\centerline{\small INFN-Roma1 and Rutgers University}
%\centerline{\small CEREMADE, Universit\'e Dauphine, Paris}
\*
\0{\bf Abstract: \it Perturbative construction of the nonequilibrium steady
  state of a rotator under a stochastic forcing while subject to torque and
  friction }
\*
\centerline{\today}
\*
\def\SEC{Introduction and models}
%%%%%%%%%%%%%%%%%%%%%%%%%%%%%%%%%%%%%%%%%%%%%%%
\section{\SEC}\iniz\label{sec1}
%\renewcommand{\theequation}{\ref{sec1}.\arabic{equation}}
%%%%%%%%%%%%%%%%%%%%%%%%%%%%%%%%%%%%%%%%%%%%%%%

Nonequilibrium stationary states can be constructed in numerical
simulations and, even for simple systems, show interesting nontrivial
properties (see for instance \cite{ILOS011}).  There are very few examples, \cite{DLS002},\cite{BDGJL01}, of systems that can be completely studied
theoretically. Even apparently very simple systems can be quite
difficult to analyze, if the aim is to describe in detail their stationary
states properties.

Here we study the stationary state of a particle, ``{\it rotator}'', bound
on a circle, in contact with a Langevin stochastic thermostat at
temperature $\beta^{-1}$ while subject to a constant torque of strenght
$\tau$.

The equation of motion is the stochastic equation on $T^1\times R$ 
\be\eqalign{\dot {q}=&\frac{{p}}J, \qquad \dot
  {p}=- {\dpr} U-\t -\frac\x{J} { p} +\sqrt{\frac{2\x}{\beta}}\dot w
},\qquad U(q)\,\defi\, 2\,g\,V\cos q\,,\Eq{e1.1}\ee
where $J$ is the inertia moment of the rotator, $\x$ is the friction $\dot
w$ is a standard white noise with increments $dw=w(t+dt)-w(t)$ of variance
$dt$ so that $\sqrt{\frac{2\x}{\beta}}\dot w$ is a Langevin random force
at inverse temperature $\b$. 

The potential $U$, Eq.\equ{e1.1}, is a smooth function on the circle $T^1$,  
with $V>0$ an energy parameter and $g$ a dimensionless parameter measuring 
the strength of the conservative force.
We chose the simplest possible function because the problem seems
difficult even in such a case,
The parameters $\b,J,V$ will be kept constant, hence the model is
  simply a pendulum subject to constant torque $\t$, friction $\x$ and
  white noise $\sqrt{\frac{2\x}\b}\dot w$.

When the torque $\t =0$, the stationary probability is simply given
by the equilibrium Gibbs state proportional to $e^{-\beta H(p,q)}dpdq$, where
$H(p,q) = \frac1{2J} p^2+ U(q)$ is the Hamiltonian of the system.
The torque $\tau$ is a non-conservative (or non-gradient) force and
drives the system out of \textit{equilibrium}.  The 
thermostatting action of the noise acting on the rotator is essential, 
as it may take energy out, allowing reaching a steady state.

The {\it overdamped case} corresponds to the limit equation obtained 
rescaling time $t'= \lambda t$ and $\xi' = \lambda\xi$ and taking the limit
$\lambda \to +\infty$: $q_\xi'(t')$ converges in law to the solution $\bar
q(t)$ of the stochastic differential equation on $T^1$
\be
    \dot {\bar q} = -\frac1\xi (\dpr U+ \t) + \sqrt{\frac{2}{\beta\xi}}
  \; \dot {\lis w}\,.
\Eq{e1.2}
\ee
The generator for a distribution $F(q)\frac{dq}{2\p}$ of such
evolution is $\LL^*_{od}F$ defined by $\LL^*_{od}F=
\x^{-1}(\dpr_q\Big((\t+\dpr_q U)F\Big)+\dpr_q^2
F)$. The process therefore has an explicit stationary
state given by
\be
  F_{od}(q) =e^{-{\beta} U(q)} 
\int_0^{2\p} e^{\beta(\t y + U(q+y))} dy / Z\,,
\Eq{e1.3}\ee
where $Z$ is the normalization (to $1$), \cite{FG012}.  In fact defining $B(y) = \tau y +
U(y)$, it is $ F_{od}(q) = Z^{-1} \int_q^{q+2\pi} e^{\beta(B(y) - B(q))}\;
dy $ and $\dpr F_{od}(q) = -\beta (\tau + \dpr U (q)) F_{od}(q)$, so that
$\LL^*_{od} F_{od}=0$.

The study of the overdamped Langevin equation on the circle $T^1$ is
 surprisingly involved, \cite{FG012}. There is indeed a remarkable lack of smoothness
 of some properties (large deviation functions) of the above stationary
 distribution in some region of the model parameters.

Going beyond the overdamped motion assumption is the aim of the present
work: hence the particle state will be described by a pair of coordinates
$(q,p)\in T^1\times R$.

The physical dimensions are $[J]=[time\times\, action]$, $[p]=[action]$,
$q=[angle]$, $[V]=[U]=[\t]=[energy]=[action/time]$, $[\x]=[action]$,
$[\b]=[energy^{-1}]$ and $\dot w$ has dimension $[\dot
  w]=[action/\sqrt{time}]$: the ``unusual'' dimensions arise because $q$ is
dimensionless (an angle) rather than with dimension of a length.

The generator for the evolution of a distribution $F(q,p)\frac{d q\,d
  p}{2\p}$ is
\be\eqalign{
\LL^*F=&-\Big\{
\Big(\frac{p}J\dpr_{q} F(q,p)
-(\dpr_{q} U(q)+\t)\dpr_{p} F(q, p)\Big)
\cr&
-\x\,\Big(\b^{-1}\dpr^2_{p}F(q,p)
+\frac1J\dpr_{p} (p\,F(q,p))\Big)\Big\}\,.\cr}
\Eq{e1.4}
\ee
The solution of $\LL^*F=0$ will be searched within the class of probability
distributions satisfying the following hypotheses:
\*

{\it \0(H1) The function $F(p,q)$ is smooth and admits an expansion in
  Hermite polynomials (or ``Wick monomials'') $H_n$ of the form
\be\eqalign{
F(q,p)=& G_{ \b}(p)\sum_a \r_a(q)\,: p^a:,\qquad 
G_\b(p)=\frac{e^{-\frac\b2 p^2}}{\sqrt{2\p\b^{-1}}}\,,\cr
:p^n:\defi&\Big(\frac{J\b^{-1}}2\Big)^{\frac{n}2}
H_n(\frac{p}{\sqrt{2J\b^{-1}}})\cr}\,,\Eq{e1.5}\ee
where $a\ge0$ are integers; so that $\int :p^n:\,:p^m:\,G_\b(p)\,dp=\d_{nm}
n!\,(J\b^{-1})^{n}$.  \\
(H2) The coefficients $\r_n(q)$ are
$C^\infty$-differentiable in $q,g$ and the $p,q,g$-derivatives of
$F$ can be computed by term by term differentiation, obtaining convergent
series.  } 
\*

It is known, \cite{MS002}, that the equation $\LL^* F(p,q)=0$ admits a unique smooth and
positive solution in $L_2(G_\b(dp)dq)$, hence in $L_1(dpdp)\cap L_2(dpdq)$,
 with $\int F dp dq=1$.

In this paper we show that \*

\0{\bf Theorem: 
\it There is a formal power series expansion in $g$ for a
  solution of the equation $\LL^* F=0$ with coefficients $\r_n^{[r]}(q)$;
  their Fourier's transforms $\r_{n,k}^{[r]}$ can be determined by a
  constructive algorithm, vanish for $|k|>r$ and satisfy the bounds
\be \x^{n}|\r_{n,k}^{[r]}|\le A_r\frac{(B_r)^{\,n}}{n!}\d_{|k|\le r} ,
\qquad \forall r, k\,, \Eq{e1.6}\ee
for $A_r,B_r$ suitably chosen, depending also on the dimensionless
parameters of the problem $\b V,\b\t,\h=\frac{\b \x^2}{2J}>0$.} \*

The properties (H1),(H2) allow us to perform the algebra needed to turn the
stationarity condition $\LL^*F=0$ into a hierarchy of equations for the
coefficients $\r_n(q)$, $\forall n\ge0$
\be \eqalign{
n& \b^{-1} \dpr\r_n(q)+\Big[\frac1J \dpr\r_{n-2}(q)+\frac\b{J} (\dpr
  U(q)+\t)\r_{n-2}(q)\cr&
+(n-1)\frac\x{J}\r_{n-1}(q)
\Big]=0\,,
\cr
}\Eq{e1.7}\ee
where $\r_{-1},\r_{-2}$ are to be set equal to zero.
\*

\0{\it Remarks:} (1) Adapting \cite{MS002}, it can be seen that $\LL^*
F(p,q)=0$ admits a unique solution smooth in $p,q$. However its analyticity
in $g$ and the properties of its representation in the form in
Eq.\equ{e1.5}, are not proven by the above Theorem, as it only deals
with the Taylor coefficients of a formal expansion in
powers of $g$.
\\(2) Identities follow immediately by normalization or by
integration. Let $\lis\r_n\defi \int \r_n(q)\frac{dq}{2\p}$ and
$\wt\r_n(q)\defi \r_n(q)-\lis\r_n$; then for $n=1$
\be\int \r_0(q)\frac{dq}{2\p}=1,\quad \wt \r_1=0\,.
\Eq{e1.8}\ee
\\ (3) The bound in Eq.\equ{e1.6} yields a convergent expression
for the $\r^{[r]}_n(q)$ or for $F^{[r]}(q,p)$ or, for bounded $f$, of $\int
F^{[r]}(q,p)$ $\cdot f(p)dp$. Such expressions can possibly be compared
with results of numerical simulations.
\*

The above Theorem is the main result. 

%%%%%%%%%%%%%%%%%%%%%%%%%%%%%%%%%%%%%%%%%%%%%
%%%%%%%%%%%%%%%%%%%%%%%%%%%%%%%%%%%%%%%%%%%%
\def\SEC{Dimensionless equations}
\section{\SEC}\label{sec2}\iniz
%%%%%%%%%%%%%%%%%%%%%%%%%%%%%%%%%%%%%%%%%%%%
%%%%%%%%%%%%%%%%%%%%%%%%%%%%%%%%%%%%%%%%%%%%
%%%%%%%%%%%%%%%%%%%%%%%%%%%%%%%%%%%%%%%%%%%%

It is convenient to introduce the dimensionless quantities 
\be
\s_n(q)\defi\r_n(q)\x^n n!,\qquad\h\defi{\b \x^2}/J, \qquad \b\t,\qquad \b V,
\Eq{e2.1}
\ee
which will be relevant in the following.

Let a tilde over a function $f(q)$ mean $f-\lis f$, with $\lis f\defi
\int_0^{2\p}f(q)\frac{dq}{2\p}\,\defi\, \media{f}$. If
$\s_n(q)\defi\wt\s_n(q)+\lis\s_n$ with $\media{\wt\s_n}=0$, consider the
sequence of functions $\s(q)=(\wt \s_n(q), \lis \s_n)_{n\ge1}$. 
The stationarity Eq.\equ{e1.7}, for $n\ge1$ and using the Hermite
polynomials properties $p\,:p^n:=p^{n+1}:+\frac{J}\b :p^{n-1}\kern-1mm:$,
$\dpr_p :p^n\kern-1mm:=n :p^{n-1}\kern-1mm:$ and $\dpr_p G(p)=-\frac\b{J}\,p
\,G(p)$, becomes, after dividing both sides by $n$
\be\eqalign{
\dpr\wt\s_{n}=&-{\h}\Big((n-1)\Big(\dpr\wt\s_{n-2}+\b \wt{{\dpr
    U\wt\s_{n-2}}} +\b \dpr
U\lis\s_{n-2}\cr
&+\b\t\wt\s_{n-2}+\wt\s_{n-1}\Big)\Big)\,,\cr
\lis\s_n=&-\Big(\lis{{\b\dpr
    U\wt\s_{n-1}}}+
 \b\t\,\lis\s_{n-1}\Big)\,,\cr}\Eq{e2.2}\ee
where $\s_{j}=0$ for $j<0$. For $n=0$, Eq.\equ{e1.7} is an
  identity, thus Eq.\equ{e2.2} holds only for $n\ge1$. By
normalization $\lis\s_0\equiv1$, as also implied by the second of
Eqs.\equ{e2.2}.

The Fourier transform of Eqs.\equ{e2.2} for $n\ge1$ is

\be\eqalign{
\wt\s_{n,k}=&-\frac{\h(n-1)}{ik}\wt\s_{n-1,k} 
-\h(n-1)(1+\frac{\b\t}{ik})\wt\s_{n-2,k}
\cr
&-\h(n-1)\Big(
\frac{\b  g V}{ik}\sum_{k'=\pm1}ik'\wt\s_{n-2,k-k'}+
\b  g V \lis\s_{n-2}\d_{k,\pm1}\Big)\,,
\cr
\lis\s_n=&-
\Big(\lis{{\b\dpr
    U\wt\s_{n-1}}}+\b\t 
\lis\s_{n-1}\Big)\cr
\equiv& (-\b\t)^n+
\sum_{j=0}^{n-1}(-\b\t)^j\lis{{(-\b\dpr U)\wt\s_{n-1-j}}}\,. 
\cr}\Eq{e2.3}\ee

After defining $\V S_{n,k}\defi\pmatrix{\wt\s_{n,k}\cr\wt\s_{n-1,k}\cr}$, it
is natural to introduce the
$g$--indepen\-dent $2\times2$ matrices $M_{n,k}$
\be
M_{n+1,k}\defi
\pmatrix{-\frac{n}{i\,k}\eta& 
in a_k\eta\cr
1 &0\cr},\qquad a_k\defi(i+\frac{\b\t}{k})\,,
%M_{n+1,k}^{-1}\defi&
%\pmatrix{0           & 1\cr
%\frac{1}{i n a_k\h}  & -\frac{1}{k\,a_k}\cr}
\Eq{e2.4}\ee 
so that Eqs.\equ{e2.3} can be written more concisely, for
$n\ge0$,
\be \eqalign{
\V S_{n+1,k}=& M_{n+1,k}\Big( \V S_{n,k}+\V X_{n+1,k}\Big),\quad 
{\V X}_{n+1,k}\defi
{0 \choose x_{n+1,k}}\,,\cr 
x_{n+1,k}\defi&-\frac{\b g V}{ia_k}\Big( \d_{|k|=1}{\lis\s}_{n-1}+
\sum_{k'=\pm1} \frac{k'}{k}\wt\s_{n-1,k-k'}\Big)\,,\cr
\noalign{\vskip2mm}
{\lis\s}_{n+1}=&-(\lis{{\b\dpr U\wt\s_{n}}}+\b\t {\lis\s}_{n})\,.\cr
}\Eq{e2.5}\ee
Notice that $M_{n,k}$ is defined for $n\ge1$, while $M_{n,k}^{-1}$ is
defined for $n\ge2$.  \*

\0{\it Remarks:} (1) The case $g=0$ is an exercise: $\s_n(q)$ must be
constant simply by symmetry, and $\s_0=1$.  It follows, from Eqs.\equ{e2.2}
and \equ{e1.4} $\s_n\equiv\lis\s_n={(-\b\t)^n}$ and the series in
Eq.\equ{e1.4} becomes
\be\eqalign{\s_n\equiv&\,\lis\s_n={(-\b\t)^n}\,,
\cr
 F(p,q)=&G_\b(p) \sum_{n=0}^\infty \frac{(-1)^n}{n!}
\Big(\frac{\b\t}{\x}\Big)^n
\Big(\frac{J\b^{-1}}2\Big)^{\frac n2} H_n(\frac
    {p}{\sqrt{(2\b^{-1}J)}})\,,\cr}\Eq{e2.6}\ee
which, by the definition of the Hermite polynomials via their generating
function (\cite[8.957]{GR965}), is
\be F(p,q)=G_\b(p-v), \qquad
  v=-\frac{J\t}\x\,.
\Eq{e2.7}
\ee

\0(2)  Eqs.\equ{e2.5} implies for $n=0,1$ (\Ie for $n=1,2$ in Eq.\equ{e2.3}) that

\be \V S_1={0\choose \wt\s_0},\quad \V S_2={\wt\s_2\choose0 }\,.\Eq{e2.8}\ee
{\it It is important to notice that Eq.\equ{e1.8} (hence also \equ{e2.3}) implies
  $\wt\s_1\equiv0$.}

If $\wt \s_2$ is known, $\wt\s_0$ can be derived by solving,
if possible, the equation (see the first of Eqs.\equ{e2.3} or \equ{e2.5})
\be \wt\s_{2,k}=i\h a_k\wt\s_{0,k}-\h\frac{\b g V}{ik}\sum_{k'=\pm1}
  ik'\wt\s_{0,k-k'}-\h\b g V\d_{|k|=1}\,.\Eq{e2.9}\ee

\0(3) Eqs.\equ{e2.5} can be regarded as equations for
$\V S_n, n\ge1$: if ${\V S}_1\equiv{0\choose\wt\s_0}$ ({\it i.e.} if
$\wt\s_0$) is known, then all $\V S_n$, $n\ge3$, can be immediately
computed. To this end, $\V S_1$ and $\lis\s_1$ are computed from
Eqs.\equ{e2.5} ($\lis\s_0=1$), together with $x_{2,k}$; then, from the
pair $(\V S_1,\lis\s_1)$ and $x_{2,k}$, we compute $\V S_2,\lis\s_2$ and $x_3$, by Eq.\equ{e2.5}, hence $\V S_3,\lis\s_3$ and $x_4$ follow \&tc.

\0(4) Eqs.\equ{e2.2} and its Fourier transform \equ{e2.3} will be
considered as equations for $\V S_{n,k}$ for $n\ge2$ to be solved under
the condition that ${\V S}_2={\wt\s_2\choose 0}$. The condition that the
second component of ${\V S}_2$ be $0$ is the only condition to
impose {\it a priori} and to use in the construction.

%%%%%%%%%%%%%%%%%%%%%%%%%%%%%%%%%%%%
%%%%%%%%%%%%%%%%%%%%%%%%%%%%%%%%%%%%
\def\SEC{Perturbation expansion}
\section{\SEC}
\label{sec3}\iniz
%%%%%%%%%%%%%%%%%%%%%%%%%%%%%%%%%%%%
%%%%%%%%%%%%%%%%%%%%%%%%%%%%%%%%%%%%

Consider the expansion in $g$ by writing 

\be
\eqalign{\wt\s_n(q)=&g \wt\s_n^{[1]}(q)+g^2
\wt\s_n^{[2]}(q)+\ldots\,,\cr
\lis\s_n=&\lis\s^{[0]}_n+g \lis\s_n^{[1]}+g^2
\lis\s_n^{[2]}+\ldots\,,\cr}\Eq{e3.1}
\ee
and, correspondingly, $F=F^{[0]}+g F^{[1]}+\ldots$\,.

Eqs.\equ{e2.5} become a recursive relation for the the real constants
$\lis\s^{[r]}_n$ and for the vectors $\V
S^{[r]}_{n,k}\defi { \wt \s^{[r]}_{n,k}\choose\wt \s^{[r]}_{n-1,k}},\, k>0$,
keeping in mind the c.c. symmetry $ \wt \s^{[r]}_{n,k}=\lis{{\wt
    \s^{[r]}_{n,-k}}}$. The order $r=0$ is simply
$\wt\s^{[0]}_n=0$, $\lis\s_0=1$ and for $r\ge0$ the recursion will be
conveniently written in matrix form for $n\ge0$
\be\eqalign{
 {\V S}^{[r]}_{n+1,k} =&M_{n+1,k} \Big({\V S}^{[r]}_{n,k}+{
   x}^{[r]}_{n+1,k}{0\choose1}\Big), \quad
\ \V S^{[r]}_{0,k}=y^{[r]}_{0,k} {1\choose0}\defi{\V Y}^{[r]}_{k}\,,\cr
\lis\s^{[r]}_{n+1}=&-\b\t \lis\s^{[r]}_{n}+v^{[r]}_{n+1},\quad
\lis\s^{[0]}_n=(-\b\t)^n,
\cr
}
\Eq{e3.2}\ee
where $y^{[r]}_{0,k}\equiv \wt\s^{[r]}_{0,k}$ (so that, by
Eq.\equ{e1.8},\equ{e2.8}, $\V S^{[r]}_{1,k}= \pmatrix{0\cr
  y_{0,k}^{[r]}\cr}$) and $v^{[r]}_n,x^{[r]}_{n,k}$ are given by
\be\eqalign{
x^{[r]}_{n+1,k}\defi&
-\frac{\b V}{ia_k}\Big(\lis\s^{[r-1]}_{n-1}\d_{|k|=1}+
\sum_{k'=\pm1}\frac{k'}k
\wt\s^{[r-1]}_{n-1,k-k'}\Big),\quad n\ge0\,,
\cr
v^{[r]}_{n+1}\defi&-\sum_{k'=\pm1}\b V\,ik'\wt\s^{[r-1]}_{n,-k'},\quad n\ge0
,\qquad \V {{\bf S}}^{[r]}_{1,k}=
\pmatrix{0\cr y_{0,k}^{[r]}\cr}\,,
\cr
x^{[r]}_{2,k}=&\frac{-\b V}{ia_k}
(\d_{|k|=1}\d_{r=1}+\sum_{k'=\pm1} \frac{k'}k \wt\s^{[r-1]}_{0,k-k'})\,,\cr
}\Eq{e3.3}\ee
and depend on quantities of order lower than $r$.
Here, quantities of order lower than $0$ are interpreted as $0$.

Therefore, at order $r$, all coefficients are determined in terms of the
constants $\wt\s_{0,k}^{[r]}\equiv y_{0,k}^{[r]}$ and of the $\V
S^{[r']}_{n,k'}$ with $r'<r$; since the only harmonics in $U(q)$ are
$k=\pm1$, it can be supposed that $\wt{\V S}^{[r]}_{n,k}\equiv 0$ for
$|k|>r$ and all $n$.

Furthermore, the harmonics $\wt\s_{0,k}^{[r]}$ and $\wt\s_{0,-k}^{[r]}$
are complex conjugate, so it is possible to always suppose $k>0$
and to interpret harmonics with negative values of $k$ as complex conjugate 
of the corresponding harmonics with $k>0$.

%%%%%%%%%%%%%%%%%%%%%%%%%%%%%%%%%%%%
%%%%%%%%%%%%%%%%%%%%%%%%%%%%%%%%%%%%
\def\SEC{The expansion coefficients}
\section{\SEC}
\label{sec4}\iniz
%%%%%%%%%%%%%%%%%%%%%%%%%%%%%%%%%%%%
%%%%%%%%%%%%%%%%%%%%%%%%%%%%%%%%%%%%

The recursion can be reduced to an iterative determination of
$x^{[r]}_{n,k}$ and $\lis\s^{[r]}_n$ starting from $r=1$, as the case $r=0$ has
been already evaluated at the end of Sec.\ref{sec2}, namely
$x^{[0]}_{n,k}=0$, $\lis\s^{[0]}_n=(-\b\t)^n$ and $\V S^{[0]}_n=0$, since
$\wt\s^{[0]}_n\equiv0$.

For $r\ge1$ it is, see Eqs.\equ{e2.9} and \equ{e3.3}, 
\be
\eqalign{\V S^{[r]}_{0}\defi&\V Y^{[r]}={y^{[r]}\choose0},\quad \V
S_{1}^{[r]}\defi{0\choose y^{[r]}}\,, \cr
\V S^{[r]}_{2}\defi&\pmatrix{i\,
  a_k\,\h \,(y^{[r]}+x^{[r]}_2)\cr0\cr},\qquad \lis\s^{[r]}_0=0\,,\cr}
\qquad r\ge1\,.\Eq{e4.1}
\ee
Eqs.\equ{e3.2}, for $r\ge1$, are related to the general
 $r$-independent equations for $n\ge2$, conveniently written by means of
the inverse matrix $M_{n+1,k}^{-1}$ as
\be \eqalign{
&M_{n+1,k}^{-1}\defi
\pmatrix{0           & 1\cr
\frac{1}{i\, n\, a_k\, \h}  & -\frac{1}{k\,a_k}\cr}\,,
\cr
\V S_n=&M_{n+1,k}^{-1}\V S_{n+1}-\V X_{n+1},\qquad 
\V S_2=y'{1\choose0}\,,\cr
\noalign{\vskip-2mm}
{\lis\s}_n=&(-\b\t) {\lis\s}_{n-1}+v_n,\quad 
{\lis\s}_{0}=w,\quad \V X_n=x_n{0\choose1},\ x_0=x_1=0\,.
\cr}\Eq{e4.2}\ee
This is inhomogeneous in the unknowns $(\V S_n,{\lis\s}_n)_{n\ge 2}$,
imagining $\V S_2,\V X_n,$ $v_n,w$ ({\it i.e.} $y\,, x_n\,, v_n\,, w$) as known
inhomogeneous quantities depending on the orders $r$ and $k$, as prescribed
by Eqs.\equ{e3.3}. In the following, Eqs.\equ{e4.2} will be considered for different orders $r$. Moreover, $w=0$ if $r\ge1$ and $w=(-\b\t)^n$ if $r=0$.

A few properties of products of matrices $M_{n,k}$ will be needed.

\0Let $(M^{-1}_p)^{*s}\defi M^{-1}_p\cdots M^{-1}_{p+s-1}$ for $s\ge1$,
$(M^{-1}_p)^{*0}\defi1$ and define
\be\eqalign{
\Bx_n\defi&-\sum_{h=n}^{\infty}
(M_{n+1}^{-1})^{*(h-n)}\V X_{h+1}\,,\cr
{\lis\s}_n\defi& \sum_{s\ge 1}^{n} (-\b\t)^{n-s}v_{s} +w\,,
%+(-\b\t)^{n}w
\cr} \qquad n\ge 2\,,\Eq{e4.3}
\ee
then $M^{-1}_{n+1} \Bx_{n+1} = \Bx_n + \V X_{n+1}$, if the series converges
and if $y'$ is left free. Eq.\equ{e4.3} is thus a special solution of the
recursion \equ{e4.2} ($n\ge2$).

To proceed, we introduce some notation; given $k$, let
\be \eqalign{
&\ket{\uparrow\,}\defi\pmatrix{1\cr0}\equiv\ket1,\qquad \ket{\downarrow}\defi
\pmatrix{0\cr1}\equiv\ket0,
%\qquad \g_{m}\defi \frac{4k^2 a_k}{\h m i}
\cr
&\L_{\n,\n'}(n,h)\defi \bra{\n} (M_{n+1}^{-1})^{*(h-n)}\ket{\n'},\qquad 
\n,\n'=0,1\,,\cr
&\L(n,h)\defi \L_{0,0}(n,h),\qquad \L(n,n)=1\equiv
\braket{\downarrow}{\downarrow}\,,\cr
&\z(n,h)\ \defi\ \frac{\bra{\uparrow}(M_{n+1}^{-1})^{*(h-n)}
\ket{\downarrow}}{\L(n,h)},\qquad \z(n,n)=0\equiv\braket{1}{0}\,. \cr
}\Eq{e4.4}\ee
Notice that, by explicitly computing the vectors $\bra{\n}M_{n+1}^{-1}$
and $M_h^{-1}\ket{\n'}$, the general relation (due to the special form, 
Eq.\equ{e4.2}, of
$M_j^{-1}$) is
\be \bra{\n}(M_{n+1}^{-1})^{*(h-n)}\ket{\n'}
=\frac{\L(n+\n,h-\n')}{(i\h a_k(h-1))^{\n'}},\quad 
\n,\n'=0,1\,,\Eq{e4.5}\ee
which, provided $\L(n,h)\ne0$, implies the identities
\be 
\eqalign{
&\Bz(n,h)
\defi\ {\z(n,h)\choose 1},\ \z(n,h)=\frac{\L(n+1,h)}{\L(n,h)}, 
\qquad 2\le n\le h\,,\cr
&
(M_{n'+1}^{-1})^{*(n-n')}\Bz(n,n')=\frac{\L(n',N)}{\L(n,N)}
\Bz(n',N),\quad n+1> n'\,.\cr}\Eq{e4.6}\ee
with $\z(n,n)=0$. Here, the second relation will be called the
{\it eigenvector property} of the $\z(n,h)$. Eq.\equ{e4.5} also implies the
recurrence
\be \eqalign{
\f(n,h)\,&\defi\,-\frac{\z(n,h)} {k
  a_k}=\frac{1}{1+\frac{z}{n}\f(n+1,h)}\cr
&=
\frac1{1+ \frac{\textstyle z}n \frac1{1+\frac{\textstyle z}{n+1}}\cdots
\frac{{}}{\frac{1}{1+\frac{\textstyle z}{h-2}}}},\qquad h-2\ge n,\qquad z\,\defi\,
\frac{a_k k^2}{i\h}\cr}\Eq{e4.7}
\ee
and $\f(n-1,n)=1,\f(n,n)=0$, representing the $\z$'s as continued
fractions and showing that $\z(n,h)$ and the limit
$\z(n,\infty)$ are analytic in $z$ for $|z|<\frac14$,
\cite[p.45]{CPVWJ008}.

The continued fraction for $\f(n,\infty)$ is the $S$-fraction
$\frac{n-1}{z}\mathop{\bf K}_{m=n-1}^\infty(\frac{z/m}1)$, following
\cite[p.35]{CPVWJ008}, and defines an holomorphic function of $z$ in the
complex plane cut along the negative real axis (see
\cite[p.47,(A)]{CPVWJ008}). The $\f(n,h)$ is also a (truncated) $S$-fraction
obtained by setting $m=\infty$ for $m\ge h-1$ in the previous continued
fraction. Hence, by \cite[p.47,(B)]{CPVWJ008}, $\f(n,h)$ has the same
holomorphy properties as $\f(n,\infty)$. Furhermore, $\f(n,h)$ is
holomorphic for $|z|<\frac14$, continuous and bounded by $\frac12$ in
$|z|\le\frac14$, \cite[p.45]{CPVWJ008}.

Relevant inequalities can be derived from the inequality
in \cite[p.138]{CPVWJ008}, see Sec.\ref{sec6}.

The definitions imply $\Bx_n\equiv
-\sum_{h=n}^{\infty}x_{h+1}\L(n,h)\Bz(n,h)$ (see Eq.\equ{e4.3}).
Furthermore, if the limits
$\lim_{N\to\infty}\frac{\L(2,N)^{-1}}{\L(n,N)^{-1}}$ exist, and are
symbolically denoted $\frac{\L(2,\infty)^{-1}}{\L(n,\infty)^{-1}}$, then
\be \V T^0_n=\frac{\L(2,\infty)^{-1}}{\L(n,\infty)^{-1}}\Bz(n),\qquad
\Bz(n)\defi\Bz(n,\infty)\Eq{e4.8}\ee
is a solution of Eq.\equ{e4.2} with $\V X=0$ and some initial data for
$n=2$.

A solution to the $r$-th order equations will 
thus have the form
\be
\V S_n = \Bx_n + \l \V T^0_n\,, \Eq{e4.9}
\ee 
where the constant $\l$ will be fixed to match the data at $n=2$, \Ie
null $\V S_2$ second component, see
Eqs.\equ{e2.8},\equ{e3.2}. Concerning the $r$-th order equation, the
initial data of interest are $\lis\s^{[r]}_1$ and
$y^{[r]}_{2,k}$ and $\V X^{[r]}_{n,k}, v^{[r]}_{n}$ are given by
Eq.\equ{e3.3}, in terms of quantities of order $r-1$.

Therefore there seems to be no freedom, because we expect
that the condition that $\V Y$ is proportional to $\ket{\uparrow}$, \Ie 
that its second component is $0$, fixes the free constant $\l$. Thus the (unique) 
solution to the recursion with initial
data $\V S^{[r]}_2=y\ket{\uparrow}$ has necessarily the form
\be \V S^{[r]}_{2,k}=
-\sum_{h=2}^\infty x^{[r]}_{h+1,k}\L(2,h)
(\Bz(2,h)-\Bz(2))
\Eq{e4.10}\ee
which is proportional to $\ket{\uparrow}$, because the second components of
$\Bz(2,h)$ and $\Bz(2)$ are $1$; hence in Eq.\equ{e4.9}
\be \l = \sum_{h=2}^\infty x^{[r]}_{h+1,k}\L(2,h)\,. \qquad \Eq{e4.11}
\ee
Proceeding formally, for $n>2$,$\V S^{[r-1]}_n$ will be given
by applying the recursion. Since $\Bx_n$ is a
formal solution and $\Bz(n)$ has the eigenvector
property Eq.\equ{e4.6}, it is
\be \eqalign{
\V S^{[r]}_{n,k}=&\sum_{h=2}^{n-1}
x^{[r]}_{h+1,k}\frac{\L(2,h)\L(n,\infty)}{\L(2,\infty)} \z(n)\cr
&-\sum_{h=n}^\infty x^{[r]}_{h+1,k}\Big(\L(n,h)\z(n,h)-
\frac{\L(2,h)\L(n,\infty)}{\L(2,\infty)}\z(n)\Big)\,.\cr}
\Eq{e4.12}\ee
It should be stressed that the series in Eq.\equ{e4.3} might diverge;
nevertheless, cancellations may (and will) occur in Eq.\equ{e4.12}, so that
Eq.\equ{e4.3} would still be a solution, if the series in Eq.\equ{e4.12} converges (as
can be checked by inserting it in the equation Eq.\equ{e4.2}).

To compute the first component of $\V S^{[r]}_{n,k}$, we left multiply
Eq.\equ{e4.12} by $\bra{\uparrow}$, considering that
$\z_k(n,m)=\frac{\L(n+1,m)}{\L(n,m)}$ by the first of the~\equ{e4.6} and
that $\L(n+1,n)= \z_k(n,n) \L(n,n) \equiv 0$. Setting $\L(n,m)=0\,,
\forall\, m<n$, we obtain
\be \eqalign{\wt\s^{[r]}_{n,k}=&\sum_{m=2}^{n}
x^{[r]}_{m+1,k}\L(2,m)\frac{\L(n,\infty)}{\L(2,\infty)}
\Bz(n)\cr 
&-\sum_{m=n+1}^\infty x^{[r]}_{m+1,k}\Big(\L(n,m)\Bz(n,m)-
\L(2,m) \frac{\L(n,\infty)}{\L(2,\infty)}\Bz(n)\Big)\cr
\equiv&
\sum_{m=2}^{n}x^{[r]}_{m+1,k}
\Big(\prod_{j=2}^{m-1}\frac{\z(j,\infty)}{\z(j,m)}\Big)
\Big(\prod_{j=m}^{n}\z(j,\infty)\Big)
\cr&-\sum_{m=n+1}^\infty x^{[r]}_{m+1,k}
\Big(\prod_{j=n+1}^{m-1}\frac1{\z(j,m)}\Big)
\Big(1-\prod_{j=2}^{n} \frac{\z(j,\infty)}{\z(j,m)}\Big),\cr}
\Eq{e4.13}\ee
for $n\ge2$. From $\wt\s^{[r]}_{2,k}$ and $x^{[r]}_{2,k}$ (derived
from Eq.\equ{e3.3}) and using $\wt\s^{[r]}_{1,k}=0$ (see
Eq.\equ{e4.1}) the ``main unknown'' $y^{[r]}_{0,k}$, {\it i.e.}
$\s^{[r]}_{0,k}$, is computed.

As stressed above, Eqs.\equ{e4.10},\equ{e4.13} are acceptable if the series
converge. When $r=1$, they do for $\b\t$ small enough using
$x^{[1]}_{n,k}=-\frac{\b V}{i a_k} (-\b\t)^{n-1}$ $\d_{|k|=1}$, since
(i) the squared norm of $M_{n+1,k}$ in $C^2$ is equal to 
\be \eqalign{
%||M_{n+1,k}||^2=
& \frac12(1+\frac1{\h^2 n^2|a_k|^2}+\frac1{k^2|a_k|^2})
%\cr&\cdot
\Big(1+
\Big({1-\frac{4 \frac1{\h^2 |a_k|^2 n^2}\frac1{k^2|a_k|^2} }
{(1+\frac1{\h^2 n^2|a_k|^2}+\frac1{k^2|a_k|^2})^2}}
\Big)^{\frac12}\Big)\cr}
\Eq{e4.14}\ee
\Ie to the square root of the maximum eigenvalue of $M_{n+1,k}^{-1\,*}M_{n+1,k}^{-1}$
and (ii) the convergence condition of the first of Eqs.\equ{e4.3} is $\b\t
||M_{2,1}^{-1}||<1$, meaning that $\b\t$ should be small enough. As it
will be seen from the estimates of Sec.\ref{sec6}, this restriction on
$\b\t$ can be removed, because of the cancellations that occur between the two addends
in Eq.\equ{e4.10}). Therefore, it will be possible to try an iterative
construction, $\forall\, \h,\b\t,r>0$.

%%%%%%%%%%%%%%%%%%%%%%%%%%%%%%%%%%%%
%%%%%%%%%%%%%%%%%%%%%%%%%%%%%%%%%%%%
\def\SEC{A constructive algorithm}
\section{\SEC}
\label{sec5}\iniz
%%%%%%%%%%%%%%%%%%%%%%%%%%%%%%%%%%%%
%%%%%%%%%%%%%%%%%%%%%%%%%%%%%%%%%%%%

If $x^{[r]}_{n,k},\lis\s^{[r]}_n$ are known, it is possible to
compute $\wt\s^{[r]}_{n,k}$ from Eqs.\equ{e3.3},\equ{e4.13} for all $n\ge0$. In
particular,
\be\eqalign{
\wt \s^{[r]}_{2,k}=&x^{[r]}_{3,k}\z(2,\infty)
-\sum_{m=3}^\infty x^{[r]}_{m+1,k}\L(3,m)\Big(1-\frac{\L(2,m)\L(3,\infty)}
{\L(2,\infty)\L(3,m)}\Big)\,,\cr
\wt\s^{[r]}_{0,k}=&\frac1{i\h a_k}\wt \s^{[r]}_{2,k}+
\frac{\b V}{ia_k}\Big(\d_{|k|=1}\d_{r=1} + \sum_{k'=\pm1}
\frac{k'}{k}\wt\s^{[r-1]}_{0,k-k'}\Big)
\Big),\qquad \wt\s_1\equiv0\cr}
\Eq{e5.1}\ee
The order $r=1$ is therefore known because $x^{[1]}_{m+1,k}=\frac{-\b
  V}{ia_k} (-\b\t)^{m+1}\d_{k,1}$, provided the series in
Eqs.\equ{e5.1},\equ{e4.13} converge. It is convenient to
introduce the kernels $\th_k(n;m)$ to abridge the Eqs.\equ{e4.13},\equ{e5.1}
into the form:
\be \eqalign{
  \wt\s^{[r]}_{n,k}=&\sum_{m=0}^{\infty}\th_k(n;m)\,x^{[r]}_{m+1,k}, \quad
  n\ne1,\quad \th_k(n;0)=0,\quad \forall n\ge 2\,,
\cr
  \th_k(n;m)=&\Big(\prod_{j=2}^{m-1} \frac{\z_k(j,\infty)}{\z_k(j,m)}\Big)
  \Big(\prod_{j=m}^{n}\z_k(j,\infty)\Big),\quad 2\le m\le n\,,
\cr
  \th_k(n;m)=&\Big(\prod_{j=n+1}^{m-1}\frac1{\z_k(j,m)}\Big)
  \Big(\prod_{j=2}^{n}\frac{\z_k(j,\infty)}{\z_k(j,m)}-1\Big), 
\quad 2\le  n<m\,,
\cr \th_k(0,m)\defi& \frac{\th_k(2,m)}{i\h a_k}\d_{m\ge2}  - 
  \d_{m,1},\qquad \th_k(1;m)\equiv0 \,,
\cr } \Eq{e5.2}\ee
where products over an empty set of labels are interpreted as $1$.

Consider the sequences $\V Z={\wt\s_{n,k}\choose\wt\s'_{n,k}}$,
$n=0,1,\ldots, \,k=\pm1,\pm2,\ldots$, in the particular cases $\V
Z^{[r]}={\wt\s^{[r]}_{n,k}\choose\wt\s^{[r-1]}_{n,k}}$, $n=0,1,\ldots,
\,k=\pm1,\pm2,\ldots$.  Remark that $\V Z^{[1]}_{n,k}=
\pmatrix{\wt\s^{[1]}_{n,k}\cr0}$ and $\wt\s^{[1]}_{n,k}$ is known (see
comment after Eq.\equ{e5.1} and remark that $\wt\s^{[0]}=0$), if the series
defining it converges. Then Eqs.\equ{e4.13},\equ{e5.2} can be rewritten as

\be{ \V Z}^{[r]} =\BB \V Z^{[r-1]},\qquad r\ge2,\qquad \V Z^{[1]}_{n,k}=
\pmatrix{\wt\s^{[1]}_{n,k}\cr0}\,,\Eq{e5.3}\ee
with the map $\BB$ defined by taking into account the above relations and also
Eqs.\equ{e3.3},\equ{e4.2}, which yield 
\be\kern-3mm\eqalign{
\lis\s^{[r]}_n=& 
\Big((-\b\t)^{n}\d_{r,0}-\b V\sum_{k'=\pm1}ik'
\sum_{h=0}^{n-1}(-\b\t)^h {{ \wt\s^{[r-1]}_{n-1-h,-k'}}}\Big),
\quad n\ge1\,,
\cr
x^{[r]}_{n,k}=&-\frac{\b V}{i
  a_k}\Big(\lis\s^{[r-1]}_{n-2}\d_{|k|=1}+
\kern-2mm\sum_{k'=\pm1}\kern-1mm
\frac{k'}{k} \wt\s^{[r-1]}_{n-2,k-k'}\Big),\qquad n\ge2\,,\cr
x^{[r]}_{2,k}=&-\frac{\b V}{ik a_k}(\d_{|k|=1}\d_{r=1}+
\sum_{k'=\pm1}k'\wt\s^{[r-1]}_{0,k-k'})\,,
\cr
}\Eq{e5.4}\ee
$\forall n\ne1,|k|\le r$.
Therefore, if $\wt \th_k(n;m) 
\defi \sum_{h=0}^\infty (-\b\t)^{h}\th_k(n,m+h+1)\,\d_{m\ge1}$ and 
the kernels $t_k(n;m),\wt t_k(n;m)$ are defined as
\be \eqalign{
t_k(n,m)\defi&\frac{-\b V}{i k a_k}\, \th_k(n;m)\, \d_{m\ge2}\,,\quad n\ge2
\cr
\wt t_k(n,m)\defi&
\frac{(\b V)^2}{a_k}\,  \wt \th_k(n;m)\,,\quad n\ge2\cr
t_k(0,m){{\buildrel \phantom{{def}}\over=}}&\frac{t_k(2,m)}{i\h
  a_k}\d_{m\ge2}+\d_{m,0}\frac{\b V}{i k a_k}\,,\quad 
\wt t_k(0,m)=\frac{\wt t_k(2,m)}{i\h a_k}\d_{m\ge1}\,,\cr
t_k(1,m){{\buildrel \phantom{{def}}\over=}}&\,\wt t_k(1,m)=0\,,\cr
}\Eq{e5.5}\ee
the map $\BB$ acquires the form ${y'_{n,k}\choose
  z'_{n,k}}=\BB{y_{n,k}\choose z_{n,k}}$, with
\be \eqalign{
y'_{n,k}=&\sum_{m=1}^\infty\sum_{k'=\pm1}
k'\Big( t_k(n,m) y_{m-1,k-k'} +\d_{|k|=1}\wt t_k(n,m) z_{m-1,-k'}\Big)\,,\cr
z'_{n,k}=&\,y_{n,k}\,,\cr}\Eq{e5.6}\ee
In conclusion, if\\ 
(i) $\V Z^{[1]}$ is given,
\\(ii) the series over $m$
  converge,
  \\ the map $\BB$ determines $\V Z^{[r]}$ for
$r\ge2$. Convergence will be studied in Sec.\ref{sec6}. \*

\0{\it Remark:} an interesting check is that, if $\t=0$, the
recursion gives $\lis\s^{[r]}_n\equiv0, \forall n\ge1,r\ge0$ and
$\wt\s^{[r]}_{2,k}=0, \forall r>0$ and this leads, as expected, to
$\s_n\equiv0,\forall n>0$ and $\s_0(q)=Z^{-1}e^{-\b\, 2\,g\,V\,\cos q}$, after
some algebra and after summation of the series in $g$.
\*

%%%%%%%%%%%%%%%%%%%%%%%%%%%%%%%%%%%%
%%%%%%%%%%%%%%%%%%%%%%%%%%%%%%%%%%%%
\def\SEC{Bounds and theorem proof}
\section{\SEC}
\label{sec6}\iniz
%%%%%%%%%%%%%%%%%%%%%%%%%%%%%%%%%%%%
%%%%%%%%%%%%%%%%%%%%%%%%%%%%%%%%%%%%
 
Let $z=\frac{k^2a_k}{i\h}\equiv (1+\frac{\b\t}{ik})\frac{k^2}\h$.  If
$z=|z|e^{2i\a}$, $|\a|<\frac12{\rm arctg}\frac{\b\t}{|k|}< \frac\p4$ (so
that $|\cos\a|>\frac1{\sqrt2}$), let $\l_{\ell,\pm}^\a\defi
\sqrt{1+\frac{4|z|}{(\ell-1)\cos^2\a}}\pm1$; then from the inequality in \cite[p.138]{CPVWJ008}
we derive that
\be
|\f(j,\infty)-\f(j,h)|\le 2\sqrt2 \prod_{\ell=j}^{h-2}
\frac{\l^\a_{\ell,-}}{\l^\a_{\ell,+}}\,\defi\,\D(j,h)\,,
\Eq{e6.1}\ee
for $h\ge j+1$ and $\f(j,j+1)=1,\f(j,j)=0$.
Furthermore, the recursion for $\f(j,h)$ implies $|{\rm
  arg}(\frac{z\f(j+1,h)}{j})|\le|{\rm arg}(\frac{z}{j})|\le 2\a$, and 
\be\eqalign{
&\Big|\f(j,h)-1\Big|\equiv|\frac{z \f(j+1,h)}{j
+z \f(j+1,h)}|\le 1\,,
\cr
&\Big|\f(j,h)^{-1}-1\Big|\equiv|\frac{z}j\f(j+1,h)|\le
2\frac{|z|}j\,,
\cr} \qquad h\ge j+2\,,
\Eq{e6.2}\ee
therefore
\be
\Big|\frac{\f(j,\infty)}{\f(j,h)}-1\Big|\le \Big(1+2\frac{|z|}j\Big)
\,\D(j,h),\qquad h\ge j+2\,.
\Eq{e6.3}\ee
These inequalities are useful to estimate the kernels $\th_k(n;n')$ appearing
in Eqs.\equ{e5.2} ({\it i.e.} in Eqs.\equ{e4.13} written as
$\wt\s^{[r]}_{n,k}=\sum_{m=0}^{\infty}\th_k(n;m)\,x^{[r]}_{m+1,k}$) and, using $\frac{\sqrt{1+x}-1}{\sqrt{1+x}+1}\le \frac{x}{1+x}$, imply
\be\eqalign{ 
\frac1{1+\frac{2\sqrt2|z|}j}\le &
\Big|\frac{\z_k(j,\infty)}{k a_k}\Big|\le 2\,,
\cr
\Big|
  \frac{\z_k(j,\infty)}{\z_k(j,m)}-1\Big|\le&\, 
\Big(1+\frac{2\sqrt2|z|}j\Big)2\sqrt2 \prod_{\ell=j}^{m-2}
\frac{\sqrt{1+\frac{8|z|}{\ell-1}}-1}
{\sqrt{1+\frac{8|z|}{\ell-1}}+1}
\cr
\le
& 
8\Big(1+\frac{|z|}{j}\Big) \prod_{\ell=j}^{m-2}\frac{8|z|}
{\ell-1+8|z|},\qquad m\ge j+1\,.
\cr}\Eq{e6.4}\ee
Hence (as $|1-X|=|e^{\log X}-1|\le e^{|\log X|}-1$) 
\be\eqalign{&\Big|1-\prod_{j=2}^n
\frac{\z_k(j,\infty)}{\z_k(j,m)}\Big|
%\cr& 
\le\Big[e^{\sum_{j=2}^n
    \log\Big(1+8\Big(1+\frac{|z|}{j}\Big)\prod_{\ell=j}^{m-2}
    \frac{8|z|}{\ell-1+8|z|}\Big)}\Big] -1\,.\cr
} \Eq{e6.5}\ee
Sacrificing better bounds for simpler ones, for $n \ge2$ and $m<n$, it is
\be\eqalign{
&|\prod_{j=2}^{m-1} \frac{\z(j,\infty)}{\z(j,m)}|
\le e^{\sum_{j=2}^{m-1} 8(1+|z|) 
  \frac{(8|z|)^{-(m-1-j)}}{(m-1-j)!}} \le e^{8(1+|z|)e^{8|z|}}\,,
\cr
&|\prod_{j=m}^n\z(j,\infty)|\le (2|k a_k|)^{n-m+1},\qquad 
|\th(n;m)|\le |2\,k\,a_k|^{n-m+1} e^{8 e^{9|z|}}\,.
\cr
}\Eq{e6.6}\ee
and, for $m\ge n$, using $e^X-1\le e^X X, \forall X\ge0$
\be\kern-3mm\eqalign{&\frac{|\th_k(n;m)|}{|k a_k|^{n-m+1}}\le 
\Big(\kern-2mm\prod_{j=n+1}^{m-1}\kern-3mm(1+\frac{4|z|}j)\Big)
\Big(e^{\sum_{j=2}^n
\Big(1+\frac{8|z|}j\Big) \prod_{\ell=j}^{m-1}
\frac{8|z|}{\ell-1+|z|}}-1\Big)
\cr
&\le e^{4 |z|(\log m-\log n)}\Big((1+8|z|) \sum_{j=2}^n 
\frac{(8|z|)^{m-j}}{(m-j)!}-1\Big)
\cr
&\le
e^{4 |z|(m-n)}
\Big(e^{(1+8|z|)\frac{(8|z|)^{m-n}}{(m-n)!}\sum_{j=2}^n
\frac{(8|z|)^{n-j}}{(n-j)!}}-1\Big)
\cr
&\le
e^{4|z|(m-n)}\Big(e^{8e^{9|z|}\frac{(8|z|)^{m-n}}{(m-n)!}}-1\Big)
\le 8 e^{9|z|} e^{8 e^{9|z|+8|z|}}
\frac{(8|z|e^{4|z|})^{m-n}}{(m-n)!}\,.
\cr}\Eq{e6.7}\ee
Therefore, $\forall z$ such that $|{\rm arg}\,z|<\frac\p4$
\be\eqalign{
\frac{|\th_k(n;m)|}{|2\,k\,a_k|^{n-m+1}}\le&\,C(z)\,
\Big(\frac{(2^4 e^{5|z|})^{m-n}}{(m-n)!}\Big)^{\d_{m> n}},\quad
%\cr
C(z)\le 
8 e^{9|z|} e^{2^4e^{2^5|z|}}\,.\cr}
\Eq{e6.8}\ee
This implies that the common bound on $\th_k(n;m)$ and $\wt \th_k(n;m)$ is
\be  \eqalign{
\le&  C(z)(2|k a_k|)^{n-m+1}\Big(\d_{n\ge
  m}+\frac{D(z)^{m-n}}{(m-n)!}
\d_{m>n}\Big)e^{D(z)}\qquad {\rm with}\cr
&z=\frac{k^2 a_k}{i\h},\qquad \frac{\b\t}{2|k a_k|}\le \frac12,\quad
D(z)\defi(2^4e^{5|z|})
\cr}
\Eq{e6.9}\ee
The case $n=0$ can be likewise checked to lead to the bounds Eq.\equ{e6.9}
with $n=0$, possibly adapting the definition of $C_1(k)$.  

The latter inequalities can be used in the estimate of
$|\wt\s^{[r]}_{n,k}|,\,r>1$ and also for $r=1$, because the ``extra term''
$\d_{|k|=1}\d_{r=1}$ in Eq.\equ{e5.4}, \Ie the only non trivial term in
$x^{[1]}_{n,k}$, does not affect the bounds, up to a redefinition of the
constants.

If $r=1$, then $|k|=1$ and, if $n\ge2$
\be\eqalign{
|\wt\s^{[1]}_{n,k}|\le&
\sum_{m\ge2} (\b\t)^m |\th_k(n,m)|\cr
\le&\frac{\b V}{|a_1|} C(z)\left[
\frac{(\frac{2|ka_k|}{\b\t})^n-1}{\frac{2|ka_k|}{\b\t}-1}
+e^{D(z)}\right] 2|k a_k|\,(\b\t)^n\cr
\le& \d_{|k|=1} C_1(k) |k a_k|^n\,,\cr}
\Eq{e6.10}\ee
with $C_1(k)$ suitably chosen, poorly bounded in $k$, but bounded
independently of the value of $\t$, as $\t$ varies in any prefixed bounded
interval (at $\h$ fixed).

The case $n=0$ can be likewise checked to lead to the bounds in Eq.\equ{e6.10}
with $n=0$, possibly adapting the definition of $C_1(k)$.

Define the kernels
\be T_{k,n,\a;\wt k,\wt n,\wt\a}= \cases{\frac{-\b V}{ik a_k}\th_k(n,\wt
  n)\d_{|\wt k-k|=1}&\ $\a=1,\ \wt\a=1$\cr \frac{(\b
    V)^2}{a_k}\wt\th_k(n,\wt n)\d_{|\wt k-k|=1}&\ $\a=1,\ \wt\a=2$\cr
  \d_{n,\wt n}\d_{k,\wt k}&\ $\a=2,\ \wt\a=1$\cr 0& \ $\a=2,\ \wt\a=2$\cr}\,\,,
\Eq{e6.11}\ee
and consider ${\wt\s^{[r]}_{n,k}\choose \wt\s^{[r-1]}_{n,k}}$,
as a vector with components
$\s^{[r]}_{n,k,1}\defi\wt\s^{[r]}_{n,k}$ and $\s^{[r]}_{n,k,2}\defi$ 
$\wt\s^{[r-1]}_{n,k}$. It is thus possible to write the general expression 
for $\s^{[r]}_{n,k,\a}$
\be\s^{[r]}_{n,k,\a}=\sum_{m;k';\a'} T_{n,m;
  k,k';\a,\a'}\s^{[r-1]}_{m,k',\a'}\,,\Eq{e6.12}\ee
and bound it by
\be \sum_{\{n_i\},\{k_i\},\{\a_i\}}\kern-5mm 
 \d_{|k_{r-1}|=1}
 \prod_{i=1}^{r-1}|T_{n_{i-1},n_{i};k_{i-1},k_{i};\a_{i-1},\a_{i}}|
|\wt\s^{[1]}_{n_{r-1},k_{r-1}}|\,,
\Eq{e6.13}\ee
with $n_0=n,k_0=k,\a_0=\a$. 

Inserting the bounds \equ{e6.9} and \equ{e6.10} into Eq.\equ{e6.13}\footnotemark\footnotetext{Remark that the summation over the labels $k_i$
involves at most $3^r$ choices (as there are only three choices for
$k_i-k_{i+1}$ in Eq.\equ{e6.13}), while the summation over the labels
$\a_i$ involves $2^r$ choices (due to the two possibilities for the labels
$\a_i$)},
% using the bounds in Eq.\equ{e6.9} # gia' detto all'inizio del periodo.
taking into account that $|k_j|\le r-j$, summing a few elementary (geometric and exponential) series, the bound $|\s^{[r]}_{n,k}|\le A_r (B_r)^n $, $r>1$, for suitable $A_r,B_r$ follows
and the theorem is proved.

%%%%%%%%%%%%%%%%%%%%%%%%%%%%%%%%%%%%
%%%%%%%%%%%%%%%%%%%%%%%%%%%%%%%%%%%%
\def\SEC{Weak small scale dissipation. Conclusions}
\section{\SEC}
\label{sec7}\iniz
%%%%%%%%%%%%%%%%%%%%%%%%%%%%%%%%%%%%
%%%%%%%%%%%%%%%%%%%%%%%%%%%%%%%%%%%%

The bounds on $|\r^{[r]}_{n,k}|$ hold with a suitable choice of the
constants $A_r,B_r$ which are positive if $\h,\b\t >0$.  Therefore a formal
asymptotic series at $g=0$ is $\forall R\ge0$:
\be\r^{[\le R]}(q,p)= \sum_{r=0}^R g^r 
\sum_{k=-R}^R\sum_{n=0}^\infty 
e^{i q k}\r_{n,k}^{[r]} :p^n:\Eq{e7.1}\ee  
is bounded by (using the bound on Hermite functions, \cite[8.954.2]{GR965})
\be \eqalign{
|\r^{[\le R]}(p,q)|\le& \sum_{r=0}^R \sum_{k=-R}^R\sum_{n=0}^\infty 
\lis A_r g^r B_r^n\frac1{n!}
\frac{\sqrt{n!} 2^{1-n/2}}{\x \sqrt{\h^n}} e^{+\frac\b{4J} p^2}\cr
|F^{[\le R]}(p,q)|\le&\, e^{-\frac\b{4J}p^2}  B'_R
\cr
}\Eq{e7.2}\ee
with a suitably chosen $B'_R$.
A natural question is whether having obtained non convergent bounds
is due to a poor estimate or to a singularity at $g=0$.

The mechanical system has two qualitatively different motion regimes: if
$\t>0$ is fixed then for $g$ small ($\t\gg g V$) the pendulum will in
the average rotate on a time scale of order $\frac{J\t}\x$; if, instead,
$g$ is fixed and $\t$ small ($\b\t\ll g\b V$), the pendulum will oscillate, 
very rarely performing full rotations. 

Our theory, however, gives a formal power series with estimates uniform in
$\b\t$, for $|\b\t|$ in a bounded interval, thus the power series would
be the same for $g$ in an interval where motions with $\t\gg g V$ or
$\t\ll gV$ take place.  Therefore we think that our series is really only
formal and resummations have to be devised that would treat differently the
large $\t$ and large $g$ regimes.  \*

\0{\it Remarks:} (1) If the series for $F$ could be shown to
converge (in ``any'' sense) it could be concluded that the coefficients
that have been formally computed are indeed the coefficients of the Hermite
expansion for $F$, as even the positivity of $F$ could be derived by the
following simple argument.  Adapting the proof in \cite{MS002}, any $0\le
f\in L_1\cap L_\infty$ function with $\int f dp dq=1$ has the property that
$\lim_{t\to+\infty} e^{\LL^* t}f=F_\infty>0$ with $\LL^*
F_\infty=0$ and $F_\infty$ is the unique smooth, positive, integrable,
normalized solution of $\LL^* F_\infty=0$. Hence if $f_+,f_-$ denote the
positive and negative parts of $F$ it will be
\be F\equiv e^{\LL^* t} F\equiv e^{\LL^* t} f_+-e^{\LL^* t}
f_-\tende{t\to+\infty}(a-a')F_\infty
\Eq{e7.3}\ee
if $\int f_+=a,\int f_-=a'$. Therefore $a-a'=1, F=F_\infty, a=1,a'=0$ and
this would prove that $F>0$ and that it is the unique stationary probability
density in $L_1\cap L_2$ for the process with generator $\LL^*$, {\it i.e.}
for the stochastic equation Eq.\equ{e1.1}.

\0(2) The algorithm in Sec.\ref{sec4} can be implemented numerically,
by computing with prefixed precision a prefixed number of coefficients
$\r^{[r]}_{n}(q)$, and programming the recursion in Eq.\equ{e5.6}.

\0(3) The overdamped case, see Sec.\ref{sec1}, can be treated in the same
way: in this case $\r_n\equiv0$ for $n\ne0$ and $\r_0(q)$ is analytic in
$g$ for $g$ small: it is possible to study the distribution $\r_0$ in
  great detail and for all $g$, obtaining interesting results on large
fluctuations of several observables: see \cite{FG012}.

\0(4) The continued fractions in Eq.\equ{e4.7} are really remarkable and a
natural question is whether they are related to known special functions or,
alternatively, which are their properties near the negative real axis.
\*

\0{\bf Acknowledgements: \it This work has been partially supported by the
  European Advanced Grant {\em Macroscopic Laws and Dynamical Systems}
  (MALADY) (ERC AdG 246953). GG is grateful to INFN-Roma1, Rutgers University
  and CNRS for support.}
\*

\0{\small\tt giovanni.gallavotti@roma1.infn.it, 
iacob@ceremade.dauphine.fr,\\
olla@ceremade.dauphine.fr}

%%%%%%%%%%%%%%%%%%%%%%%%%%%%%%%%%%%%%%%%%%%%%%%%%%%%%%%%%%
\def\SEC{References}
%%%%%%%%%%%%%%%%%%%%%%%%%%%%%%%%%%%%%%%%%%%%%%%%%%%%%%%%%%
%%%%%%%%%%%%%%%%%%%%%%%%%%%%%%%%%%%%%%%%%%%%%%%%%%%%%%%%%%

\bibliographystyle{unsrt}
%\bibliography{0Bib}

\begin{thebibliography}{1}

\bibitem{ILOS011}
A.~Iacobucci, F.~Legoll, S.~Olla, and G.~Stoltz.
\newblock Negative thermal conductivity of chains of rotors with mechanical
  forcing.
\newblock {\em Physical Review E}, 84:061108 +6, 2011.

\bibitem{DLS002}
B.~Derrida, J.~L. Lebowitz, and E.~R. Speer.
\newblock Exact free energy functional for a driven diffusive open stationary
  nonequilibrium system.
\newblock {\em Physical Review Letters}, 89:030601, 2002.

\bibitem{BDGJL01}
L.~Bertini, A.~De Sole, D.~Gabrielli, G.~Jona-Lasinio, and C.~Landim.
\newblock Fluctuations in stationary nonequilibrium states of irreversible
  processes.
\newblock {\em Physical Review Letters}, 87:040601, 2001.

\bibitem{FG012}
A.~Faggionato and G.~Gabrielli.
\newblock A representation formula for large deviations rate functionals of
  invariant measures on the one dimensional torus.
\newblock {\em Annales de l' Institut H. Poincar\'e, (Probabilit\'e et
  Statistique)}, 48:212--234, 2012.

\bibitem{MS002}
J.~C. Mattingly and A.~M. Stuart.
\newblock Geometric ergodicity of some hypo-elliptic diffusions for particle
  motions.
\newblock {\em Markov Processes and Related Fields}, 8:199--214, 2002.

\bibitem{GR965}
I.S. Gradshtein and I.M. Ryzhik.
\newblock {\em Table of integrals, series, and products}.
\newblock Academic Press, New York, 1965.

\bibitem{CPVWJ008}
A.~Cuyt, V.~Petersen, B.~Verdonk, H.~Waadeland, and W.~Jones.
\newblock {\em {Handbook of Continued Fractions for Special Functions}}.
\newblock Springer, Berlin, 2004.

\end{thebibliography}

\end{document}